# Analysis of Marketed versus Not-marketed Mobile App Releases

Maleknaz Nayebi
SEDS laboratory
University of Calgary
Calgary, Canada
mnayebi@ucalgary.ca

Homayoon Farrahi
SEDS laboratory
University of Calgary
Calgary, Canada
homayoon.farrahi@ucalgary.ca

Guenther Ruhe
SEDS laboratory
University of Calgary
Calgary, Canada
ruhe@ucalgary.ca

## ABSTRACT

Market and user characteristics of mobile apps make their release managements different from proprietary software products and web services. Despite the wealth of information regarding users' feedback of an app, an in-depth analysis of app releases is difficult due to the inconsistency and uncertainty of the information. To better understand and potentially improve app release processes, we analyze major, minor and patch releases for releases following semantic versioning. In particular, we were interested in finding out the difference between marketed and not-marketed releases. Our results show that, in general, major, minor and patch releases have significant differences in the release cycle duration, nature and change velocity. We also observed that there is a significant difference between marketed and non-marketed mobile app releases in terms of cycle duration, nature and the extent of changes, and the number of opened and closed issues.

## 1. INTRODUCTION

There is an increasing trend towards more adaptive delivery of software products. This trend includes more flexibility in the release cycle duration with a tendency for continuous integration and delivery in order to respond more quickly to user needs. Mobile apps are special types of software products having commonalities and differences to traditional products. In a former study, Nayebi et al. [3] asked developers about their rational for their app release decisions. In this paper, we performed a more in depth study of Android mobile apps. We looked into the evolution of major, minor and patch releases. In particular, we look into releases using semantic versioning from open source apps. To better understand the status quo, we analyzed the key WH's questions, they are related to: Where, When, What, and Which of app releases.

Starting from open source apps in F-Droid, we filtered out apps that did not follow semantic versioning and analyzed the number and cycle of releases (*When?*), nature of changes (*What?*), reported issues, change requests (*Why?*), and extent and domain of changes (*Which?*). Our investigations showed an incompatibility between marketed and not-marketed releases. We called the releases being in both GitHub and app store as *marketed releases*. However, we found releases in GitHub that were not published in the app store, these are not-marketed releases.

We also found that some releases never found their way to the end users. We call them (*Not-marketed releases*). To better understand the rational for releasing an app version to the market, we also compared the not-marketed releases with the marketed ones. More specifically, the following research questions are studied in this paper:

**RQ1:** How major, minor and patch releases are different in terms of their release cycle duration, quality, amount of change and the nature of changes?

**RQ2:** What are the characteristics of app releases being released (marketed) or not getting released (not-marketed)?

## 2. DATA COLLECTION

To analyze the development and release process of apps, we gathered information from F-droid [1], a well-known repository of open source apps. At the time that this analysis was conducted, F-Droid listed 1,844 Android apps. With the aim of an in depth analysis of *major*, *minor* and *patch* releases, we filtered out the apps which were not following *semantic versioning* [4]. Semantic versioning or Semver [4] is a versioning schema with three digits in the version name and a format of MAJOR.MINOR.PATCH. In this versioning system, an increment to the MAJOR digit should be made when incompatible API changes happen, MINOR increments should happen when functionality changes in a backwards compatible manner, and PATCH increments are applied for backwards-compatible bug fixes [4][5]. Among all the apps in F-Droid, 69 apps follow semantic versioning. In this paper, we focus on these 69 apps to answer our two research questions.

## 3. DESCRIPTIVE AND INFERENCE ANALYSIS

In this section, we first answer RQ1 related to all of the 69 apps and then investigate RQ2 by comparing the marketed and not-marketed releases.

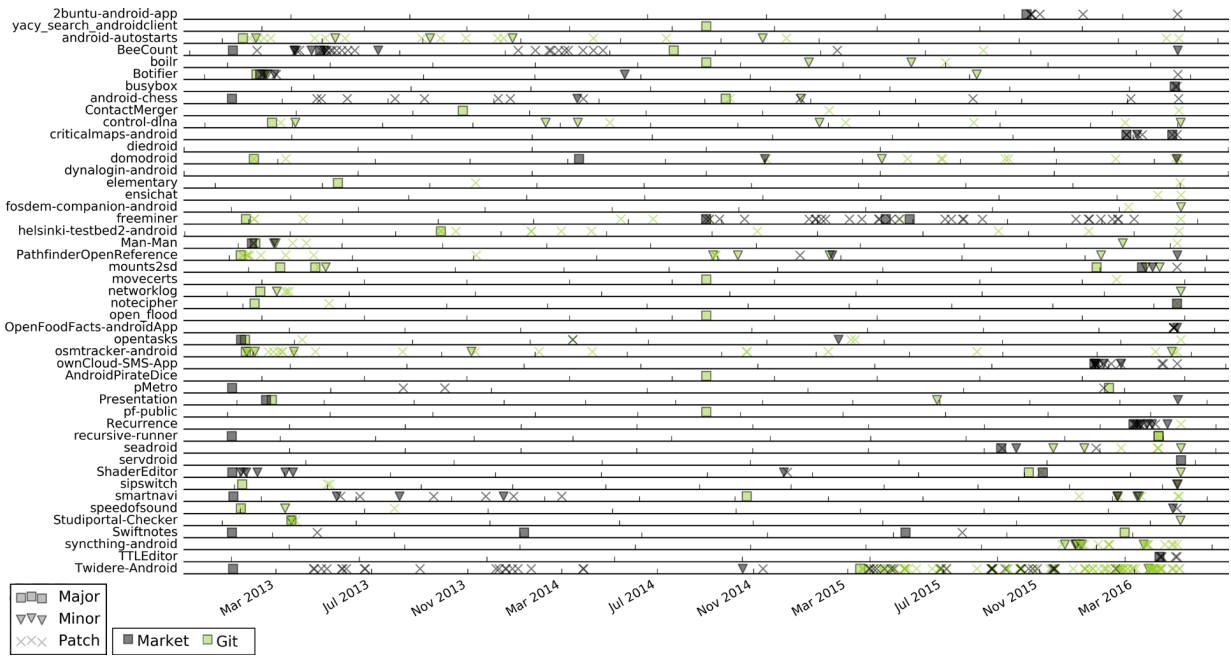

**Figure 1: Delta versions of major, minor and patch releases for apps released in GooglePlay XOR in GitHub.**

### 3.1 Repositories & categories - Where?

Looking at *all* the F-droid apps, 73 different source control systems are used. Our 69 semantic versioning apps are hosted across *three* different repositories. The largest portion of them (91.4%) are hosted on GitHub.com. 5.7% of projects are hosted on `code.google.com` and 2.9% are hosted on `sourceforge.com`.

These apps are classified into 19 categories within Google-Play. The largest number of apps (26.1%) is in the *tools* category, 17.45% are in *music*, 11.6% in *productivity*, 11.6% in *communication. transport, social, travel* categories each have 4.3% of the apps and *finance and books* each have 2.9% of apps. *Multimedia, games, weather, simulation, news, media, lifestyle, connectivity, and education* are the rest of categories with only one app in each category. We analyzed both GitHub and GooglePlay releases of all apps and separated them based on:

**Location:** Being in GitHub only, GooglePlay only or in both (without inconsistency).

**Type** Being a major, minor or a patch release.

While analyzing the location and type of releases, we found that 22 (out of 69) F-droid apps have never been released into GooglePlay. We further investigated the compatibility between app releases in GitHub and GooglePlay and found that only two apps had *no* inconsistencies. For the remaining 47 apps, Figure 1 shows the difference (delta performance) in release dates between the two repositories. In Figure 1 the shape demonstrates the release type and the color demonstrates the location of the release. For the period between March 2013 and March 2016, there were among 990 releases across GitHub and GooglePlay, 40.7% were in both GitHub and GooglePlay. Interestingly, 23.4% were in the market but not in GitHub (called *marketed*) and 35.9% of releases existed in GitHub but were not available in GooglePlay (called *not-marketed*). We compared marketed and not-marketed releases (Figure 2) and found that:

- Significantly more patches were released in Google-Play compared to GitHub (Mann-Whitney P-value of 0.024). The analogous tests for major and minor releases were not significant.

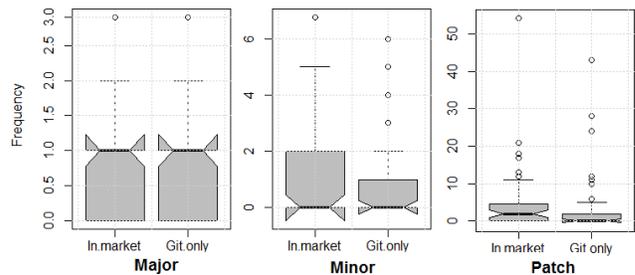

**Figure 2: Major, minor and patch releases for marketed and not-marketed releases.**

### 3.2 Release type & number of release - When?

We observed that 84.2% of apps have only *one* major release and only 5.8% of apps have more than two major releases. When it comes to minor releases, 27.1% of the apps have only one minor release and 25.3% of them have two to four minor releases. Most apps (55.2%) have less than three patches. We found:

- There is a significant correlation between the number of major and minor releases (p-value= 0.03). However, we did not find a significant correlation between patches and major or minor release types.
- Release duration of patches is significantly shorter than the one of minor releases and the duration of minor

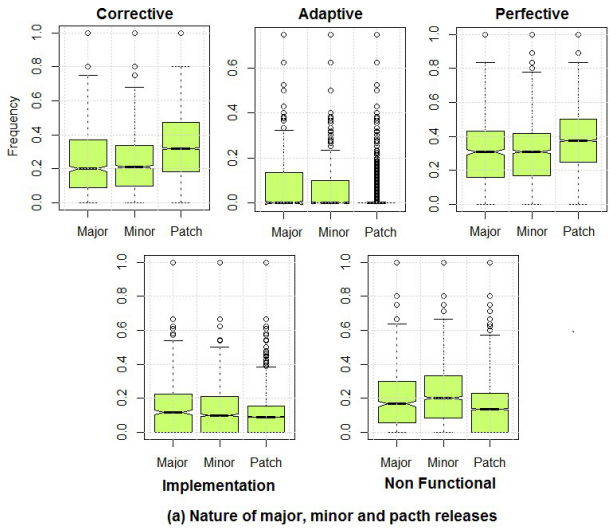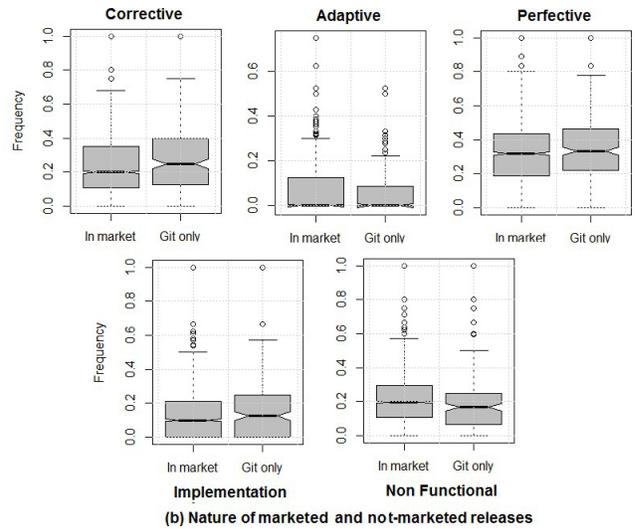

**Figure 3: Nature of changes for (a) major, minor and patch release of apps, and (b) for marketed and not-marketed apps**

releases is significantly shorter than the one of major releases.
- Distribution of release duration (in days), demonstrates the significant difference in the median of duration for major, minor and patch releases (see Figure 4).

We also compared the duration of release cycles between the marketed and not-marketed releases in Figure 4.
- We found a significant difference between release cycle duration for the marketed major versions compared to the not-marketed ones (Mann-Whitney test p-value= 0.015). Also, not-marketed patches have significantly shorter intervals in comparison to major and minor not-marketed releases (Mann-Whitney test p-value= 0.001).

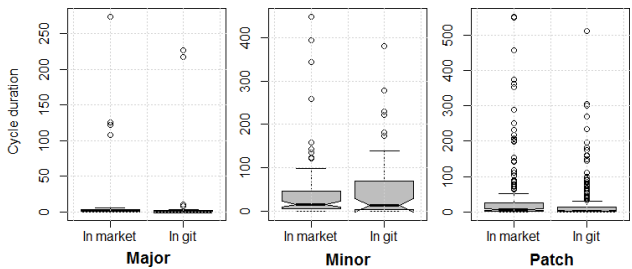

**Figure 4: Release duration for major, minor and patch releases.**

### 3.3 Nature of changes - What?

We analyzed 29,860 commit messages across all the apps based on the categories defined by Hindle et al. [2]. In this way, we considered five categories, being *corrective* (processing, performance or implementation failure), *adaptive* (change in data or processing environment), *perfective* (maintainability and performance enhancement), *implementation* (new requirements), and *non-functional* (source control management and code clean-up) [2]. Each commit was classified in at least one of the above categories (multiple classification possible). The process was done manually and independently by three software engineering students. Looking into the nature of commits for different release types, applying Kruskal-Wallis and subsequent post-hoc tests showed that:
- Patches have a significantly higher number of *corrective* (p-value= 0.001) and *perfective* (p-value= 0.001) changes.
- Adaptive, implementation and non-functional changes were significantly different between major, minor and patch releases (all p-values = 0.001).

The nature of changes for different release types is demonstrated in Figure 3-(a). Comparing the nature of marketed and not-marketed releases (Figure 3-(b)) showed that:
- Not-marketed releases have a significantly higher number of corrective (Mann-Whitney p-value= 0.002) and implementation (Mann-Whitney p-value= 0.001) commits.
- Marketed releases have a significantly higher number of non-functional commits in comparison to not-marketed releases (Mann-Whitney p-value= 0.001).

### 3.4 Extent of changes - Which?

For each release, we investigate on *churn* (lines added and deleted from code), *number of changed files* in conjunction with the release type. For this analysis, we calculated *file change velocity (# of changed files/ release cycle duration)* and *change velocity (code churn/ release cycle duration)*. Our investigations showed that:
- Churn velocity is significantly different between major, minor and patch releases (p-value= 0.001).
- File change velocity is significantly higher for major releases in comparison to minor and patch releases (Mann-Whitney p-values= 0.001). However, we could not find any significant difference in terms of file change velocity between minor and patch releases.

We also compared marketed and not-marketed releases in terms of the change extent and found that:

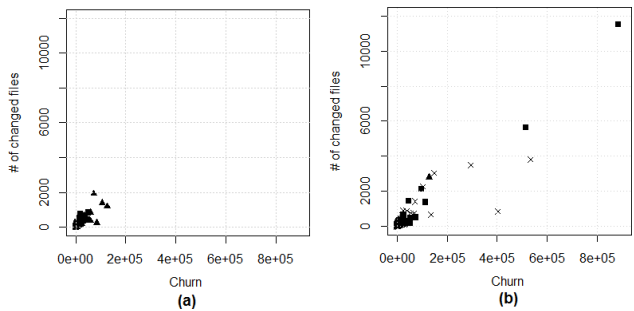
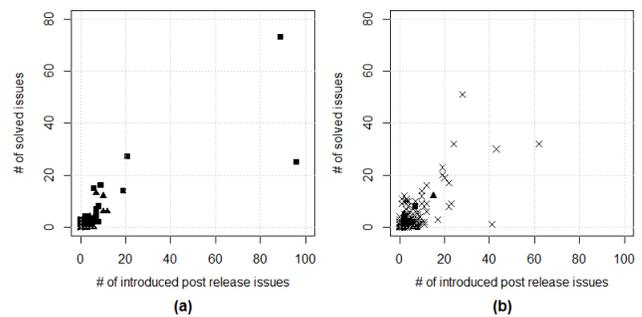

Figure 5: Extend of change and type of release for (a) marketed and (b) not-marketed release. Legend is the same as in Figure 1.

Figure 6: Number of introduced and solved post release issues for (a) marketed and (b) not-marketed release. Legend is the same as in Figure 1.

- In-market and out-market releases have a significant difference in terms of churn velocity and file churn velocity (Mann-Whitney p-values= 0.001). We found that the average of both velocities are higher for not-marketed in comparison to marketed releases.

### 3.5 Issues introduced vs issues solved - Why?

To find reasons for change velocity, we also looked into the velocity of introduced post-release issues (# of introduced post release issues/release cycle duration) and the velocity of solved issues (# of closed issues in a release/release cycle duration) for different release types (see Figure 6). We found that:

- Issues are mostly introduced and solved within very short release cycles (see Figure 7).
- Patches have a significantly higher velocity of opened (p-value = 0.0153) and solved issues (p-value= 0.0413) in comparison to the minor releases. We have not found any significant difference between other groups.
- Marketed releases have a significantly higher velocity in opening and closing issues compared to the not-marketed ones (p-value = 0.001).

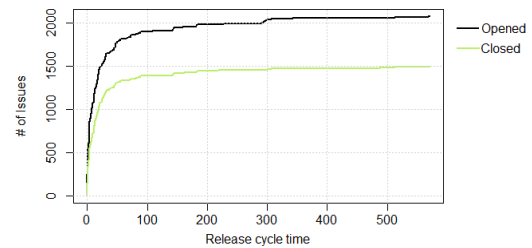

Figure 7: Cumulative chart of opened and closed issues within release cycle time (in days).

## 4. THREATS TO VALIDITY

Analysis of 69 apps just from F-droid cannot be claimed to be representative. The smaller sample size is a consequence of not following semantic versioning. To mitigate the risk of mis-classification of commit messages, we manually analyzed 29,860 messages. While of substantial effort, we believe that this helped to improve the correctness of classification.

## 5. CONCLUSIONS

There is rich literature about analytical software engineering [6-33]. We compared semantic versions of 69 Android mobile apps to: (i) find differences between major, minor and patch re- leases (RQ1) and (ii) characterize marketed and not-marketed releases (RQ2). For RQ1, we demonstrated that the release duration for patches was shorter than the one for minor and major releases. We observed that patches have more correc- tive and adaptive changes. For RQ2, we found that signif- icantly more patches are released into the market and that market releases have shorter release cycle duration. Looking into commit messages of these releases, it appears that many of them are testing a version by releasing it to just some customers or localizing their app updates in short release cycles. Our study showed that, not-marketed releases have a higher number of corrective and implementation releases while marketed releases have more non-functional releases. Also, velocity of change and opening and resolving issues are higher for marketed releases. We see the exploratory analysis results as valuable input for release engineers to objectively decide which releases should be marketed and which ones not.

## 6. ACKNOWLEDGEMENT

This research was partially supported by the Natural Sciences and Engineering Research Council of Canada, NSERC Discovery Grant 250343-12.